\documentstyle[preprint,eqsecnum,aps,epsfig]{revtex}

\tighten

\def\Journal#1#2#3#4{{#1} {\bf #2}, #3 (#4)}

\def\PRL{\em Phys. Rev. Lett.}

\begin{document}

\draft

\title{Parity Violating Measurements of Neutron Densities}

\author{C. J. Horowitz\footnote{email: charlie@iucf.indiana.edu}}
\address{Dept. of Physics and Nuclear Theory Center \\
Indiana University \\
Bloomington, Indiana 47405 USA }

\author{S. J. Pollock\footnote{email:  Steven.Pollock@colorado.edu}}
\address{Dept. of Physics , CB 390 \\
University  of Colorado\\
Boulder, CO 80309 USA }

\author{P. A. Souder\footnote{email:  souder@suhep.phy.syr.edu}}
\address{Dept. of Physics \\
Syracuse University\\
Syracuse, N.Y., USA}

\author{R. Michaels\footnote{email: rom@jlab.org}}
\address{Thomas Jefferson National Accelerator Facility \\
Newport News, VA, USA}

\date{\today}

\maketitle

\date{\today}

\newpage


\begin{abstract}
Parity violating electron nucleus scattering is a clean
and powerful tool for measuring the spatial distributions of 
neutrons in nuclei with unprecedented accuracy.
Parity violation arises from the interference of electromagnetic
and weak neutral amplitudes, and the $Z^0$ of the 
Standard Model couples primarily to neutrons at low $Q^2$.
The data can be interpreted with
as much confidence as electromagnetic scattering.
After briefly reviewing the present 
theoretical and experimental knowledge
of neutron densities, we discuss possible parity violation measurements,
their theoretical interpretation, and applications.
The experiments are feasible at existing facilities.  
We show that theoretical corrections are either small or
well understood, which makes the interpretation clean.  
The quantitative relationship to atomic parity 
nonconservation observables is examined, and we
show that the electron scattering asymmetries
can be directly applied to atomic PNC because the observables
have approximately the same dependence on nuclear shape.
\end{abstract}

\vskip 1 in


\section{Introduction}

\par The size of a heavy nucleus is one of its most basic properties.  However,
because of a neutron skin of uncertain thickness, the size does not follow from
measured charge radii and is relatively poorly known.  For example, the root
mean square neutron radius in $^{208}$Pb, $R_n$ is thought to be about 0.25 fm
larger then the proton radius $R_p\approx 5.45$ fm.  An accurate measurement of
$R_n$ would provide the first clean observation of the neutron skin.  This is
thought to be an important feature of all heavy nuclei.

The interior baryon density of a heavy nucleus is closely related to its size.
The saturation density of nuclear matter $\rho_0$ is a fundamental concept
central to nuclear structure, the nature of the interactions between nucleons,
models of heavy ion collisions and applications of dense matter in Astrophysics.
The value of $\rho_0$ is inferred from the central density of heavy nuclei, most
notably $^{208}$Pb.  One then corrects for the effects of surface tension and
Coulomb interactions (which tend to cancel) and deduces the saturation density
of an infinite system.

However, present estimates of $\rho_0$ are based only on the known proton
density.  Thus $\rho_0$ is uncertain because we do not have accurate information
on the central neutron density.  An accurate measurement of the neutron radius
$R_n$ will constrain the average interior neutron density and help refine our
knowledge of $\rho_0$.

Ground state charge densities have been determined from elastic electron
scattering, see for example ref.\cite{Sick}.  Because the densities are both
accurate and model independent they have had a great and lasting impact on
nuclear physics.  They are, quite literally, our modern picture of the nucleus.

In this paper we discuss future parity violating measurements of neutron densities.
These purely electro-weak experiments follow in the same tradition and can be
both accurate and model independent.  
Neutron density measurements may have implications for nuclear structure, atomic
parity nonconservation (PNC) experiments, isovector interactions, the structure
of neutron rich radioactive beams, and neutron rich matter in astrophysics.  It
is remarkable that a single measurement has so many applications in atomic,
nuclear and astrophysics.

Donnelly, Dubach and Sick\cite{donnelly} suggested that parity violating
electron scattering can measure neutron densities.  This is because the
$Z-$boson couples primarily to the neutron at low $Q^2$.  Therefore one can
deduce the weak-charge density and the closely related neutron density from
measurements of the parity-violating asymmetry in polarized elastic scattering.
This is similar to how the charge and proton densities are deduced from
unpolarized cross sections.

Of course the parity violating asymmetry is very small, of order a part per
million.  Therefore measurements were very difficult.  However, a great deal of
experimental progress has been made since the Donnelly {\it et. al.}  
suggestion, and since the early SLAC experiment\cite{SLAC}. 
This includes the Bates $^{12}$C experiment\cite{carbon12}, 
Mainz $^{9}$Be experiment\cite{Heil}, SAMPLE\cite{sample1} 
and HAPPEX\cite{happex1}.  The relative speed of the HAPPEX result and the very 
good helicity correlated beam properties of CEBAF show that very accurate 
parity violation measurements are possible.  Parity violation is now an 
established and powerful tool.

For example, the HAPPEX result suggests that strange quarks do not make large
contributions to the nucleon's electric form factor.  Clearly additional
experiments should (and will) be done to further measure strange quarks.
However, it is important to also apply parity violation to other physics
objectives such as neutron densities.  This will allow one to take maximum
advantage of parity violation.

It is important to test the Standard Model at low energies with atomic PNC, see
for example the Colorado measurement in Cs\cite{Wieman98,Wieman99}.  These
experiments can be sensitive to new parity violating interactions such as
additional heavy $Z-$bosons.  Furthermore, by comparing atomic PNC to higher
$Q^2$ measurements, for example at the $Z$\ pole, one can study the momentum 
dependence of Standard model radiative corrections.  However, as the 
accuracy of atomic PNC experiments improves they
will require increasingly precise information on neutron
densities\cite{Pollock,Chen}.  
This is because the parity violating interaction is
proportional to the overlap between electrons and neutrons.  In the future the
most precise low energy Standard Model test may involve the combination of an
atomic PNC measurement and parity violating electron scattering to constrain the
neutron density.

Unfortunately, atomic PNC suffers from atomic theory uncertainties in the
electron density at the nucleus.  This motivates future atomic experiments
involving isotope ratios where the atomic theory cancels.  However, these ratios
may require even more nuclear structure information on isotope differences of
neutron densities.  Parity violating electron scattering measurements of isotope
differences is beyond the scope of this paper.  Instead we focus on simpler
measurements of the neutron density in a single closed (sub)shell isotope.  These
measurements should provide an important first step for later work on isotope
differences.

There have been many measurements of neutron densities with strongly interacting
probes such as pion or proton elastic scattering, see for example ref.
\cite{Ray}.  We discuss some of these in section II.  Unfortunately, all such
measurements suffer from potentially serious theoretical systematic errors.  As
a result no hadronic measurement of neutron densities has been generally
accepted by the field.  Because of the uncertain systematic errors, modern mean
field interactions are typically fit without using any neutron density
information, see for example refs.\cite{Furnstahl,ring}.

An electro-weak measurement of the neutron density in a nucleus such as
$^{208}$Pb may allow the calibration of strongly interacting probes.  By
requiring that the hadronic reaction theory reproduce the electro-weak
measurement one should reduce theoretical errors.  This is analogous to using
beta decay to calibrate $(p,n)$ probes of Gamow Teller strength.  Once proton
nucleus elastic scattering is calibrated it should be possible to study neutron
densities in a variety of other nuclei including radioactive beams.

Finally, there is an interesting complementarity between neutron radius
measurements in a finite nucleus and measurements of the neutron radius of a
neutron star.  Both provide information on the equation of state of dense
matter.  In a nucleus, $R_n$ is sensitive to the surface symmetry energy or the
symmetry energy at low densities while the neutron star radius depends on the
symmetry energy at high densities.

In the future we expect a number of improving radius measurements for nearby
isolated neutron stars.  For example, from the measured luminosity and surface
temperature one can deduce an effective surface area and radius from
thermodynamics.  Candidate stars include Geminga\cite{Geminga} and 
RX J185635-3754\cite{Nature}.

This paper is organized as follows.  In section II we discuss the present
theoretical and experimental knowledge of neutron densities.  In section III we
present general considerations for neutron density measurements and include some
experimental issues in section IV.  Section V discusses many possible
theoretical corrections and shows that the interpretation of a measurement is
very clean.  The relationship between neutron density measurements and atomic
parity nonconservation experiments is discussed in section VI.  Finally we
conclude in section VII.

\section{Present Knowledge of Neutron Densities}

In this section we discuss our present knowledge of neutron densities. 
Unfortunately, neutron density uncertainties have not been extensively
discussed in the literature.  Fortson et al.\cite{Fortson} give some 
discussion on the present uncertainties in neutron densities and how this 
uncertainty impacts atomic PNC.  They claim a relatively large error in the 
neutron radius $R_n$ of order 10\%.  

We believe the most accurate information comes from theory.  As we discuss
below, mean field models predict a relatively small spread in neutron densities
once the effective interaction is constrained to reproduce observed charge
densities and binding energies.  We also discuss neutron density measurements
with elastic magnetic electron scattering and strongly interacting probes.

\subsection{Neutron Density Theory}

\begin{figure}
\vbox to 3in{\vss\hbox to 8in{\hss {\includegraphics{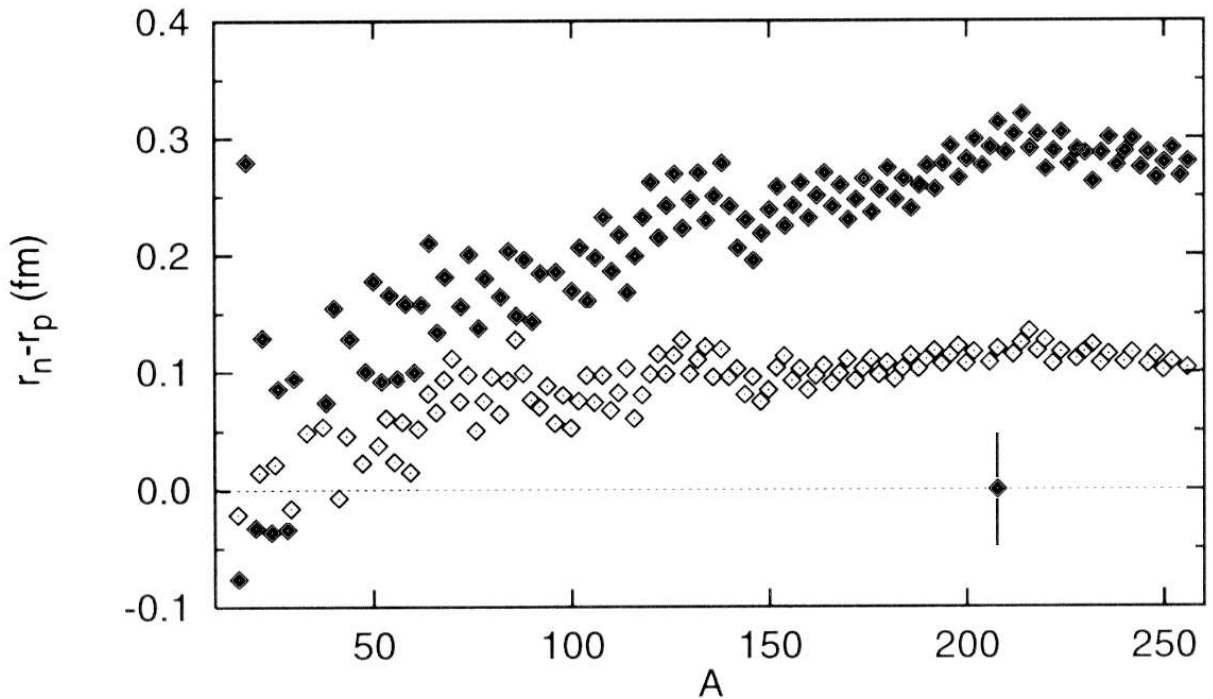}}\hss}} 
\caption{
The difference between neutron radii $R_n=r_n$ and proton radii 
$R_p=r_p$ for several nuclei of different mass number $A$.  The filled symbols
are for the relativistic mean field NL1 interaction while the open symbols
are for the nonrelativistic zero range Skyrme skiii interaction.  This figure
is taken from calculations of Ring {\it et.al} {\protect\cite{ring}}.  
A possible 1\% measurement in
$^{208}$Pb is indicated by the error bar which has been arbitrarily placed
at $R_n-R_p=0$.}
\label{rnrp}
\end{figure}

Mean field models have been very successful at reproducing many features of
nuclear charge densities including measured charge radii.  
Figure ~\ref{rnrp}, adapted
from Ring et al.\cite{ring}, shows differences between neutron $R_n$ and proton
radii $R_p$ for a range of nuclei for calculations based on two typical
interactions.  A nonrelativistic zero range Skyrme force gives $R_n-R_p\approx
0.1$Fm for $^{208}$Pb while a relativistic mean field calculation gives
$R_n-R_p\approx 0.3$Fm.  We don't claim that these two calculations represent
extreme values.  Rather they represent the range in $R_n-R_p$ for two typical
classes of interactions.

In order for the theoretical error to be less then the spread in 
Figure ~\ref{rnrp} one
must demonstrate that one (or both) of the calculations is unrealistic.  While
this may be possible in the future, it has not yet been demonstrated since
both calculations are in common use.  Of course, the real uncertainty
could be {\it larger} then the spread in Figure ~\ref{rnrp}
 if the two calculations do
not probe the full parameter space of possible mean field interactions.
In the absence of more precise uncertainties, one can take this spread 
$\approx 0.2$Fm as some measure of the present uncertainty in $R_n$.
This compares to a charge radius of 5.51 Fm.  A one percent measurement of 
$R_n$ with an accuracy of about $0.05$Fm can
clearly distinguish between the two forces.  Furthermore, it can
distinguish either prediction for $R_n-R_p$ from zero thus cleanly observing
the neutron skin.

We note that the relativistic mean field calculation with its larger $R_n$
predicts a significantly smaller central neutron density then does the Skyrme
interaction.  What gives rise to the differences between the two calculations?
Unfortunately, there is not much discussion in the literature.  We speculate that
some of the difference arises because the relativistic mean field calculation
uses a finite range force while the Skyrme interaction is zero range.  There is
an attractive interaction between the neutrons and the protons.  A finite range
force allows the neutrons to sit at a slightly larger radius and still feel much
of the attraction.  Therefore we expect finite range interactions to predict
slightly larger $R_n-R_p$ then for zero range forces.

Once mean field interactions are constrained to reproduce a neutron radius
measurement in a stable nucleus such as $^{208}$Pb they can make improved
predictions for a variety of unstable nuclei.  We note that the nonrelativistic
Skyrme force SKX\cite{brown} is designed for use in both normal and exotic 
nuclei such as $^{48}$Ni, $^{68}$Ni and $^{100}$Sn.  One example of relativistic
mean field calculations for neutron rich nuclei is ref.\cite{ring2}.  The
structure of exotic nuclei is important for Astrophysics and for radioactive 
beams.


\subsection{Neutron Density Measurements}

There have been many measurements sensitive to neutron densities.  Originally
neutron radii were extracted from Coulomb energy differences\cite{Nolen}.
However, it is now thought these measurements are sensitive to isospin violating
interactions.  Next $(p,d)$ and $(d,t)$ stripping reactions are sensitive to the
tail in the neutron density at very large radius\cite{Schiffer}.  However,
stripping reactions are not directly sensitive to the interior density.  Because
the interior density is much larger then that in the tail it contributes
significantly to $R_n$.  Therefore $R_n$ can not be extracted from stripping
experiments without making model assumptions.

Proton nucleus elastic scattering is sensitive to both the surface and
interior neutron density\cite{Ray}.  Typically this data is analyzed in 
an impulse approximation where a nucleon-nucleon interaction is folded with
the nucleon density.  Unfortunately, there are corrections to the impulse
approximation from for example multiple scattering and medium modifications
to the NN interaction whose uncertainties are difficult to quantify.  
Limitations in the theoretical analysis can show up as an unphysical dependence
of the extracted neutron density on the beam energy.

Future work on extracting neutron densities from proton scattering would
be very useful.  This could take advantage of advances in full folding
calculations.  Furthermore, neutron-nucleus elastic scattering data would be
very helpful.  Comparing proton- and neutron-nucleus scattering could help 
constrain the effective proton-neutron interaction.  Finally, if proton-nucleus
scattering can be calibrated to accurately reproduce a neutron density
measurement in a stable nucleus then it could be applied to a wide variety
of other nuclei.  For example, it is possible to measure proton-nucleus 
scattering from radioactive beams with a hydrogen target in inverse kinematics.

Finally, data comparing the elastic scattering of positive and negative
pions from nuclei exist\cite{Olmer}, but again there are uncertainties in the
analysis\cite{Pollock}.  These methods are not really directly sensitive to the
neutron density.

Elastic magnetic electron scattering is an established tool for nuclear
structure.  Furthermore, magnetic scattering is directly sensitive to the
neutron magnetic moment.  Thus information about valence neutron radii can be
extracted.  Note, more calculations of the effects of Coulomb distortions on
magnetic scattering from heavy nuclei would be useful.  See for example
\cite{wright}.  However most of the neutrons in a heavy nucleus are coupled to
spin zero and make no contribution to the magnetization.  Therefore, magnetic
scattering can not directly determine $R_n$.

We conclude that no existing measurement of neutron densities
or radii has an established accuracy of one percent.  While some conflicting
claims may have been made, all hadronic probes of $R_n$ suffer from some
reaction mechanism uncertainties.  As a result there is no agreement in 
the community that any measurement has the requisite accuracy.   Even if it
is possible to reach one percent accuracy with a hadronic probe, this 
accuracy has not yet been established.

\section{General Considerations}

In this section we illustrate how parity violating electron scattering
measures the neutron density.  For simplicity, this section uses the plane-wave
Born approximation and neglects nucleon form factors.  The effects of Coulomb
distortions and form factors are included in section V.  These
are necessary for a quantitative analysis but they do not invalidate the
simple qualitative picture presented here.

The electron interacts with a nucleus by exchanging either
a photon or a $Z_0$ boson.  The propagator involved in the
interaction is of the form
\begin{equation}
\frac{1}{Q^2+M_B^2}
\label{equation_propagator}
\end{equation}
where $M_B$ is the mass of the exchanged boson.  For the photon $M_B = 0$, 
whereas for the $Z_0$ the mass term dominates.
Since for elastic scattering from nuclei, $M_Z^2\gg Q^2$, the photon
term is much larger than the $Z_0$ term.  Note, we use the convention $Q^2=-{q_\mu}^2>0$.

Another difference between the exchange of the photon and the $Z_0$
is the couplings to both the electron and the nucleons.  The photon
has purely vector couplings, and couples only to protons at $Q^2 = 0$.  
We note that for
the spinless nuclei considered here, the magnetic moments cannot contribute.
The $Z_0$ has both vector and axial vector couplings.  Since the
nuclei being considered are spinless, the net axial coupling to the nucleus
is absent.  In contrast to the case for photons, the $Z_0$ has a much larger
coupling to the neutron than the proton.  In addition, the $Z_0$ has
a large axial coupling to the
electron that results in a parity-violating amplitude.
The effect of the parity-violating part of the
weak interaction may be isolated by
measuring the parity-violating asymmetry
\begin{equation}A_{LR}=\frac{\sigma_R-\sigma_L}{\sigma_R+\sigma_L},
\end{equation}
where $\sigma_{L(R)}$ is the cross section for the scattering
of left(right) handed electrons.  In contrast to the cross section,
the asymmetry is sensitive to the distribution of the neutrons in the nucleus.
The $Z_0$ also has a vector coupling to the electron, but this term is
neglected because the contribution cannot be isolated from the dominant
photon amplitude.

The implication of the above is that the
potential between an electron and a nucleus to a good approximation
may be written
\begin{equation}\hat V(r)=V(r)+\gamma_5A(r)\end{equation}
where the usual electromagnetic vector potential is
\begin{equation}V(r)=\int d^3r\prime Z\rho(r\prime)/|\vec r - \vec r\, \prime |\end{equation}
and where the charge density $\rho(r)$ is
closely related to the point proton density $\rho_p(r)$ given by
\begin{equation}Z\rho_p(r)=\sum_p\langle\psi_p^{\dagger}(r)\psi_p(r)\rangle.\end{equation}

The axial potential $A(r)$ depends also on the neutron density:
\begin{equation}N\rho_n(r)=\sum_p\langle\psi_n^{\dagger}(r)\psi_n(r)\rangle.\end{equation}
It is given by
\begin{equation}A(r)={G_F\over 2^{3/2}}[(1-4{\rm sin}^2\theta_W)Z\rho_p(r)-N\rho_n(r)]\end{equation}
The axial potential has two important features:
\begin{enumerate}
\item	It is much smaller than the vector potential, so
it is best observed by measuring parity violation.  It is of order one eV
while $V(r)$ is of order MeV.
\item	Since $\sin^2\theta_W\sim0.23,\ (1-4\sin^2\theta_W)$ is small and
$A(r)$ depends mainly on the neutron distribution $\rho_n(r)$.
\end{enumerate}

The electromagnetic cross section for scattering electrons with 
momentum transfer $q=(Q^2)^{1/2}$ is given by
\begin{equation}\frac{d\sigma}{d\Omega}=\frac{d\sigma}{d\Omega}_{\rm Mott}|F_p(Q^2)|^2\end{equation}
where 
\begin{equation}F_p(Q^2)=\frac{1}{4\pi}\int d^3r j_0(qr)\rho_p(r)\end{equation}
is the form factor for protons, where $j_0$ is the zero'th
spherical Bessel function.
From $F_p(Q^2)$, one may determine $R_p$.  
One can also define a form factor for neutrons
\begin{equation}F_n(Q^2)=\frac{1}{4\pi}\int d^3r j_0(qr)\rho_n(r)\end{equation}
Thus $R_n$ may be determined if $F_n(Q^2)$ is known.

\parindent=20pt

In Born approximation the parity-violating asymmetry involves the interference
between $V(r)$ and $A(r)$.  It is,
\begin{equation}A_{LR}=\frac{G_FQ^2}{4\pi\alpha\sqrt{2}}
\Biggl[4\sin^2\theta_W-1+\frac{F_n(Q^2)}{F_p(Q^2)}\Biggr]
\label{equation_bornasy}
\end{equation} The
asymmetry is proportional to $Q^2/M_Z^2$ (since $G_F\propto M_Z^{-2}$) which is
just the ratio of the propagators of Eq.~\ref{equation_propagator}.
Since 1-4sin$^2\theta_W$ is small
and $F_p(Q^2)$ is known we see that $A_{LR}$ directly measures $F_n(Q^2)$.  
Therefore, $A_{LR}$ provides a practical method to cleanly measure the
neutron form factor and hence $R_n$.


\section{Experimental Issues}

\par The experimental techniques for measuring small 
asymmetries of order 1 ppm have been successfully deployed in
parity experiments at electron
scattering facilities
\cite{SLAC}-\cite{happex1}.
The following general considerations applies to these experiments.
\hskip 0.1in 1) \hskip 0.05in
Often a compromise must be chosen between optimizing 
the parity violating signal and the signal to noise ratio.
The asymmetry generally increases with $Q^2$ while the cross
section decreases, which leads to an optimum choice of kinematics.
\hskip 0.1in 2) \hskip 0.05in
A major challenge for these measurements is to
maintain systematic errors associated with helicity
reversal at the $\approx 10^{-8}$ level.
There must be at least one, and preferably several,
methods to reverse the helicity.  Many reversals are needed
during an experiment, and they should follow a rapid and
random sequence to avoid any correlation with noise.
The helicity reversals should be uncoupled to other
parameters which affect the cross section.
Experiments must measure the sensitivity of the
cross section to these parameters, as well as the
helicity correlated differences in them.
\hskip 0.1in 3) \hskip 0.05in 
Electronic pickup of the helicity correlated signals
can cause a false asymmetry, as can helicity correlated deadtime.
\hskip 0.1in 4) \hskip 0.05in
In a count rate limited experiment in which the
detected particles must be integrated in order to get
the desired accuracy in a reasonable time, the linearity
of the detection system and the susceptibility to 
backgrounds are important issues.
For the high-rate experiments considered here, the radiation
hardness of the detectors is also an issue.
\hskip 0.1in 5) \hskip 0.05in
The beam polarization must be measured with high precision.
Online monitoring is possible using a Compton
polarimeter which is cross-calibrated using 
M{\o}ller and Mott polarimeters whose absolute 
calibrations may be $\approx$1\%.

\subsection{Choices of Target and Kinematics}

\par There are two nuclei which are of interest for
a measurement of the neutron radius to 1\% accuracy, 
${}^{208}$Pb and ${}^{138}$Ba.
They are equally accessible experimentally.
Pb has the advantage that it has the largest known
splitting to the first excited state (2.6 MeV) of any heavy
nucleus, and thus lends itself well to the use of a
flux integration technique.  Also Pb has been
very well studied, and with its simple structure 
is a good first test case for nuclear theory.
Ba has the advantage that it is one of the nuclei
being used for an atomic physics test of the
Standard Model.

\par
\begin{figure}
\vbox to 5.0in{\vss\hbox to 8in{\hss
{\includegraphics{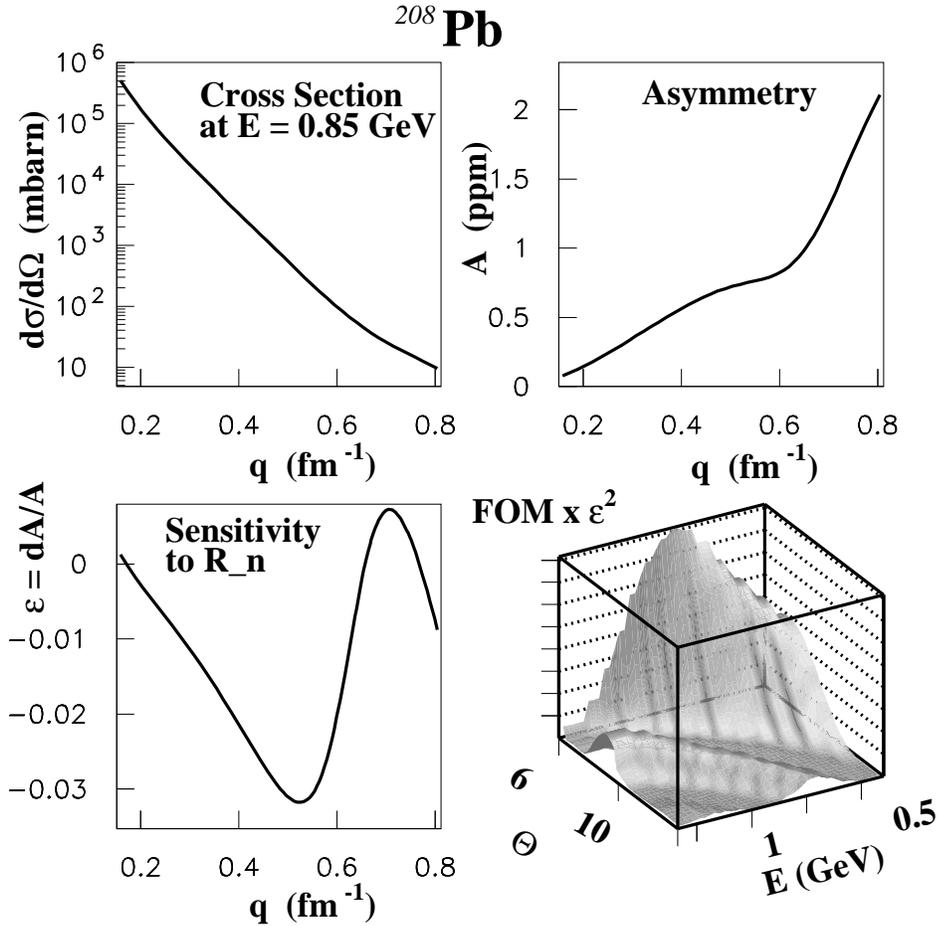}}\hss}}
\nobreak
\caption{Cross section, parity violating asymmetry, and
sensitivity to $R_n$ for ${}^{208}$Pb
elastic scattering at 0.85 GeV.  
The fourth plot shows the variation of FOM$\times {\epsilon}^2$
with energy and angle, showing an optimum at 0.85 GeV
for a $6^{\circ}$ scattering angle which corresponds to 
$q \hskip 0.03in = \hskip 0.03in 0.45 \hskip 0.03in {\rm fm}^{-1}$.}
\label{pbfom}
\end{figure}

 The choice of kinematics for a first measurement
is guided by the objective of minimizing the running
time required for a 1\% accuracy in $R_n$.  
Figure ~\ref{pbfom} shows for the case of ${}^{208}$Pb the three ingredients 
which enter into this optimization: \hskip 0.05in
the cross section $d\sigma/d\Omega$, the parity violating asymmetry $A$, and 
the sensitivity to the neutron radius $\epsilon = dA/A = (A1 - A) / A$
where $A$ is the asymmetry computed from a mean field
theory (MFT) calculation\cite{cjh} and $A1$ is the asymmetry for the
MFT calculation in which the neutron radius is increased by 1\%.
These three ingredients, which each vary with energy and angle,
are plotted in figure ~\ref{pbfom} for a beam energy of 0.85 GeV.
As we will show below, 0.85 GeV turns out to be the energy
which minimizes the running time for a 1\% $R_n$ determination.
Using magnetic spectrometers with high resolution
to isolate elastically scattered electrons, 
the optimal kinematics can be determined from 
the allowable settings for angle and momentum of the 
spectrometer by searching for the 
the minimum running time, which is equivalent to 
maximizing the product 
\begin{equation}
{\rm FOM} \hskip 0.05in \times \hskip 0.05in {\epsilon}^2
\hskip 0.05in = \hskip 0.05in R \times A^2 \times {\epsilon}^2
\end{equation}
where R is the detected rate and is proportional to $d\sigma/d\Omega$,
and ``FOM'' is the conventionally defined figure of merit for
parity experiments, FOM \hskip 0.05in$= R \times A^2$.
Note that rather than only maximizing the conventional FOM, 
parity violating neutron density measurements take into account 
the sensitivity ($\epsilon$) to $R_n$ which varies with kinematics.

\par
\begin{figure}
\vbox to 5.in{\vss\hbox to 8in{\hss
{\includegraphics{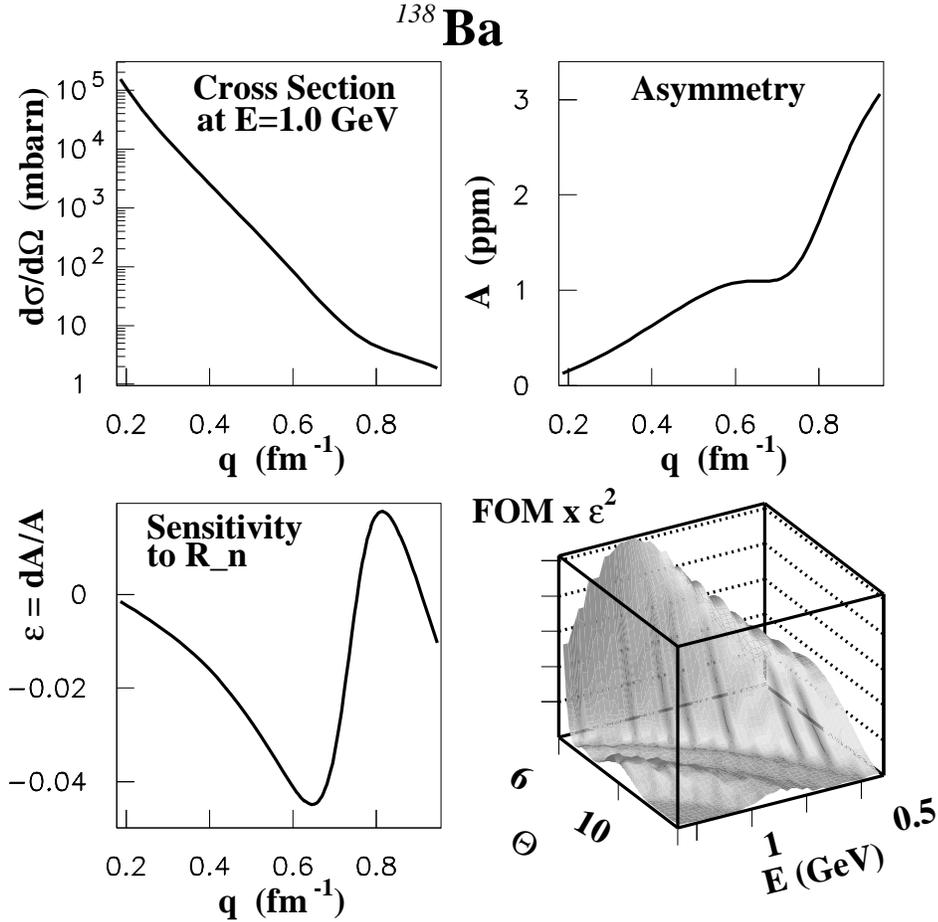}}\hss}}
\nobreak
\caption{Same as figure ~\ref{pbfom} except for ${}^{138}$Ba at 1.0 GeV}
\label{bafom}
\end{figure}

 As an example, we have performed the optimization
for the Jefferson Lab Hall A high resolution spectrometers
supplemented by septum magnets that allow to reach 
$6^{\circ}$ scattering angle.
The calculations take into account the 
averaging over the finite acceptance
and the energy resolution
needed to discriminate inelastic levels.
Figure ~\ref{pbfom} shows the product 
${\rm FOM} \hskip 0.05in \times \hskip 0.05in {\epsilon}^2$ for ${}^{208}$Pb
which peaks at E = 0.85 GeV.
Similar calculations for ${}^{138}$Ba shows 
an optimum at 1.0 GeV (figure ~\ref{bafom}).
For both these nuclei the running time $T$ in days to reach
a 1\% accuracy in $R_n$ is approximately
$T \approx 7 /( P^2 I \Omega ) $ days, where
$P$ is the polarization ($P \approx 0.8$ is achievable),
$I$ is the average beam current in $\mu$A ($I \approx 50 \mu$A is achievable)
and $\Omega$ is the solid angle acceptance
of the spectrometer in steradians.
This optimum point corresponds to 
$q = 0.45 \hskip 0.05in {\rm fm}^{-1}$
and 0.53 $\hskip 0.05in {\rm fm}^{-1}$ for
Pb and Ba respectively.
In the plots of ${\rm FOM} \times {\epsilon}^2$ one can see
a secondary ridge where one might want to perform a second
measurement at higher $Q^2$ to check the shape dependence.
Here the experimental running time becomes longer but the
required accuracy in $R_n$ can be reduced.
As an example, for ${}^{208}$Pb at E = 1.3 GeV, 
\hskip 0.05in $\theta = 8^{\circ}$,
corresponding to $q = 0.92 \hskip 0.05in {\rm fm}^{-1}$ the
running time to reach 2\% in $R_n$ 
is $T \approx 19 /( P^2 I \Omega ) $ days.

\par To reduce the running time, a thick target is needed; \hskip 0.05in
the main issues are:
\hskip 0.1in 1) \hskip 0.05in
For a given energy resolution required to discriminate
excited states, there is an optimum target thickness 
($\approx$ 10\% radiation length)
that maximizes the rate in the detector.  
As the target becomes thicker the radiative losses decrease the rate.
\hskip 0.1in 2) \hskip 0.05in
If at the low-$Q^2$ where the experiments run the rates from some low level 
inelastic states are sufficiently small and understood theoretically, 
one may tolerate accepting them into the detector, thus allowing
to integrate more of the radiative tail, typically up to 4 MeV.
\hskip 0.1in 3) \hskip 0.05in
To improve the heat load capability of the target, one may
use various ``cooling agents'', such as laminations of diamond
interleaved with the target material.  
One must have sufficient knowledge of the effect on the parity
signal.  For example, if one accepts 2\% rate from ${}^{12}$C
and the theory is understood to 1\%, the systematic error is only
0.02\%.  The theoretical error is discussed in section V.

\section{Corrections to the Asymmetry}

In this section we document a number of corrections to the parity violating
asymmetry and show that they have small uncertainties.  Therefore the
interpretation of a measurement should be clean.  We consider coulomb
distortions, strangeness and the neutron electric form factor, parity
admixtures, dispersion corrections, meson exchange currents, shape dependence,
isospin admixtures, role of excited states and the effect of target impurities.

\subsection{Coulomb distortions}

By far the largest known correction to the asymmetry comes from coulomb
distortions.  By coulomb distortions we mean repeated electromagnetic
interactions with the nucleus remaining in its ground state.  All of the $Z$
protons in a nucleus can contribute coherently so distortion corrections are
expected to be of order $Z\alpha/\pi$.  This is 20 \% for ${}^{208}$Pb.

Distortion corrections have been accurately calculated in ref.\cite{cjh}.  Here
the Dirac equation was numerically solved for an electron moving in a coulomb
and axial-vector weak potentials.  From the phase shifts, all of the elastic
scattering observables including the asymmetry can be calculated.

There are many checks on the numerics of this calculation.  First, known cross
sections including those at large angles are reproduced.  Second, the code
reproduces known plane wave asymmetries.  Finally, the sensitivity to the
subtraction between helicities is checked by varying the strength of the weak
potential.  We note that the forward angle asymmetry is much easier to calculate
then the backward angle cross section because the cross section involves extreme
cancellations in the sum over partial waves.  It is expected that the numerical
accuracy in the asymmetry is significantly better then 1\%.  However, the code
neglects terms involving the electron mass over the beam energy.  These are of
order 0.1\%.  There are now a number of independent codes which calculate the effects
of coulomb distortions\cite{ring2,Cooper,Bunny} and verify the accuracy of 
ref.\cite{cjh}.

In summary, distortion corrections are larger then the experimental error.
Furthermore, they modify the sensitivity to the neutron radius.  However, they
have been calculated with an accuracy significantly better then the expected 3\%
experimental error.

Finally, since the charge density is known it should be possible to ``invert"
the coulomb distortions and deduce from the measured asymmetry the value of a
Born approximation equivalent weak form factor at the momentum transfer $Q^2$ of
the experiment.  Thus, the main result of the measurement is the weak form
factor $F_W(Q^2)$ which is the Fourier transform of the weak charge density
$\rho_W(r)$, 
\begin{equation}F_W(Q^2) =\int d^3r j_0(qr) \rho_W(r).
\label{equation_fwq}
\end{equation}
This can be
directly compared to mean field or other theoretical calculations.  Note, $F_W$
will not be determined by comparing plane wave calculations to data.  Instead,
for example, a range of model weak densities could be adjusted until full
distorted wave calculations reproduce the experimental asymmetry.  
Then, Eq.~\ref{equation_fwq}
is used to calculate $F_W(Q^2)$.  In principle this procedure is slightly model
dependent because full distorted wave calculations need some information on $F_W(Q^2)$ for $q$ different from the single measurement.  However this model dependence is expected to be very small and can be explored by studying a variety of model densities.

\subsection{Strangeness and neutron electric form factors}

From $F_W$ the root mean square radius of the weak charge distribution $R_W$
can be determined, see section VF.  The weak radius, in turn, can be related
to the radius of the neutron distribution after making appropriate corrections.
We emphasize that the experiment measures a well defined form factor of the weak
charge distribution and that this can be directly compared to mean field models
without any additional corrections.  However, if one wishes to go further and
extract a point neutron radius one must correct for possible strange quark
contributions and other nucleon form factors.  We discuss these here.  In
addition, there could be meson exchange currents which we discuss in a later
subsection.

The electric form factors for the coupling of a $Z^0$ to the proton $G_p^Z$ and
neutron $G_n^Z$ are, 
\begin{equation}G_p^Z = {1\over 4}(G_p-G_n) - {\rm sin}^2\Theta G_p
-{1\over 4}G_s, 
\label{equation_gpz}
\end{equation} 
\begin{equation}G_n^Z = {1\over 4}(G_n-G_p) - {\rm sin}^2\Theta
G_n -{1\over 4}G_s. 
\label{equation_gnz}
\end{equation} 
We are only interested in electric form
factors since magnetic form factors make no contribution for a spin zero target.
Therefore we omit the label $E$ for clarity.  The proton (neutron)
electromagnetic form factor is $G_p$ ($G_n$).  The strange quark form factor is
$G_s$ and this is assumed the same for neutrons and protons.

We fold these form factors with point proton $\rho_p$ and neutron $\rho_n$
densities to obtain the weak charge density $\rho_W$, 
\begin{equation}
\rho_W(r)=4\int d^3r'
\bigl[G_n^Z(r')N\rho_n(|{\bf r - r'} |) + G_p^Z(r')Z\rho_p(|{\bf r -
r'}|)\bigr].
\label{equation_rhow}
\end{equation}
The densities are normalized, 
\begin{equation}
\int d^3r \rho_p(r) =
1,
\label{equation_rhop}
\end{equation}
\begin{equation}
\int d^3r \rho_n(r) = 1,
\label{equation_rhon}
\end{equation}
 and 
\begin{equation}
\int d^3r \rho_W(r)= Q_W,
\label{equation_wr}
\end{equation}
where the weak charge of the nucleus is, 
\begin{equation}
Q_W=-N + (1-4{\rm sin}^2\Theta)Z.
\label{equation_qw}
\end{equation}
The proton $R_p$, neutron $R_n$ and weak $R_W$ radii
are defined, 
\begin{equation}
R^2_p=\int d^3r r^2 \rho_p(r),
\label{equation_r2p}
\end{equation}
\begin{equation}
R_n^2=\int d^3r r^2 \rho_n(r),
\label{equation_r2n}
\end{equation}
and 
\begin{equation}
R_W^2={1\over Q_W} \int d^3r r^2 \rho_W(r).
\label{equation_rw2}
\end{equation}
It is a simple matter to calculate the weak radius from
Eq.~\ref{equation_rhow}, 


$$
-Q_W R_W^2 \hskip 0.1in = \hskip 0.1in 
N R_n^2 + (4{\rm sin}^2\Theta-1)Z R_p^2+
\bigl[N+(4{\rm sin}^2\Theta -1)Z\bigr]r_p^2 
\hskip 1.2in
$$
\begin{equation}
\hskip 1in + \hskip 0.1in \bigl[Z + (4{\rm
sin}^2\Theta-1)N\bigr]r_n^2 \hskip 0.1in + \hskip 0.1in (N+Z)r_s^2.  
\label{equation_qwrw2}
\end{equation}
Here $r_p^2$ is the mean square charge radius of the proton, $r_n^2$ 
the square of the neutron charge radius and $r_s^2$ is the mean square 
strangeness radius.

Assuming the neutron radius is much larger then $R_n-R_p$ and $r_p$ the above
reduces to


$$ R_W\approx R_n + {Z(1-4{\rm sin}^2\Theta)\over N + (4{\rm
sin}^2\Theta-1)Z} (R_n-R_p) {\hskip 2in} $$
\begin{equation}
{\hskip 1.2in} +{1\over 2 R_n} \bigl\{r_p^2
+{Z + (4{\rm sin}^2\Theta-1)N \over N+(4{\rm sin}^2\Theta-1)Z}r_n^2 + {N+Z\over
N+(4{\rm sin}^2\Theta-1)Z} r_s^2\bigr\}.
\label{equation_rwapprox}
\end{equation}

For ${}^{208}$Pb, assuming
$R_n\approx 5.50$ fm and sin$^2\Theta=0.23$, we have 
\begin{equation}
R_W\approx R_n + 0.055
(R_n-R_p) + 0.061(\pm 0.002) -0.0089(\pm 0.0003) -0.011\rho_s
\label{equation_rnapprox}
\end{equation}
in fm.  
The 0.061 is from the charge radius of the proton and the -0.0089 from the
charge radius of the neutron\cite{povh}.

The last term in Eq.~\ref{equation_rnapprox}
is from strange quark contributions.  The strange
quark form factor $G_s$ has been parameterized with $\rho_s$,
\begin{equation}
G_s(Q^2)=\rho_s\tau /(1+4.97\tau)^2,
\label{equation_gs}
\end{equation}
and $\tau=Q^2/4M^2$.  We
aim to measure $R_W$ to 1\% or about 0.055 fm.  Therefore, strange quarks will
contribute less then 1\% as long as, 
\begin{equation}
|\rho_s| < 5.
\label{equation_rhos}
\end{equation}
This is a
very mild requirement which is already strongly supported by existing
experiments.  For example, the HAPPEX measurement \cite{happex1},
\begin{equation}
G_E^s + 0.39 G_M^s (Q^2=0.48 {\rm GeV}^2) = 0.023 \pm 0.043,
\end{equation}
and the SAMPLE measurement \cite{sample1},
\begin{equation}
G_M^s(Q^2=0.1 {\rm GeV}^2) = 0.23 \pm 0.44,
\end{equation}
yield,
\begin{equation}
0.011\rho_s = -0.0043 \pm 0.021 {\rm fm}.
\label{present_limit}
\end{equation}
The errors quoted are combined statistical and systematic and we have 
assumed the form of Eq.~\ref{equation_gs} for the $Q^2$ dependence of the
strange form factor.  Equation \ref{present_limit} limits the strangeness
contributions to $R_n$ to  0.4\%. 
Note, if one assumes a different $Q^2$ dependence 
than Eq.~\ref{equation_gs},
it may be possible to somewhat weaken this limit.  However, additional
measurements in the near future will significantly tighten the constraints on
strange quarks and clearly rule out $|\rho_s|>5$.

Likewise, the neutron electric form factor contributes far less then 1\% to
$R_W$.  Theoretical models have $R_n-R_p\le 0.3$ fm, so the second term in Eq.~\ref{equation_rwapprox} is also less then 1\%.  Indeed to 1\%, the neutron radius directly follows
from the measured weak radius, 
\begin{equation}
R_n \approx R_W -0.06 {\rm fm}
\label{equation_rnapprox2}
\end{equation}
We conclude that the contribution of strange quarks 
or the neutron electric form
factor are not issues for a neutron radius measurement.  
The radius of the neutron density of a heavy nucleus can
be accurately determined from the measured weak radius.

\subsection{Parity Admixtures}

The spin zero ground state of ${}^{208}$Pb need not be a parity eigenstate.
There is probably some small admixture of $0^-$.  However, so long as the
initial and final states are spin zero, this parity admixture can not produce a
parity violating asymmetry in Born approximation\cite{spin0}.  A multipole
decomposition of the virtual photon has a $0^+$ coulomb but no $0^-$ multipole.
So long as the exchanged virtual photon is spin zero, there is no parity violating
interference because there is only a single operator.  This statement is true
regardless of the parity of the initial or final states or if the photon
coupling involves a parity violating meson exchange current.  Therefore, parity
admixtures should not be an issue for elastic scattering from a spin zero
nucleus.

\subsection{Meson Exchange Currents}

Meson exchange currents MEC can involve parity violating meson couplings.  These
are not expected to be important for a spin zero target, see the subsection on
parity admixtures above.  Meson exchange currents could also change the
distribution of weak charge in a nucleus.  However, mesons are only expected to
carry weak charge over a distance much smaller then $R_n$.  This should not lead
to a significant change in the extracted neutron radius.  Let $r_{MEC}^2$ be the
square of the average distance weak charge is moved by MEC.  Then following 
Eq.~\ref{equation_rwapprox} the 
correction to the weak radius will be of order $r_{MEC}^2/R_n$.  This
is expected to be very small because $R_n$ is large.

\begin{figure}
\vbox to 2.in{\vss\hbox to 8in{\hss {\includegraphics{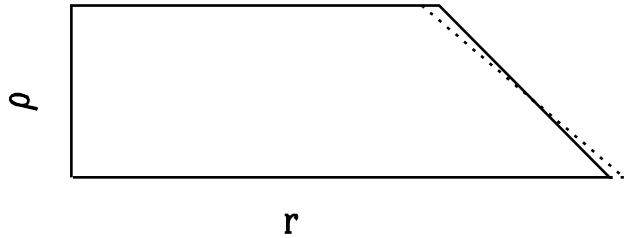}}\hss}} 
\caption{
The weak charge density of a heavy nucleus (schematic).  Meson
exchange currents and or nucleon form factors can only change the density in the
surface region.  This is indicated by the dotted line.  The density in the
interior is insensitive to MEC.}
\label{mec}
\end{figure}

This same result can be viewed another way.  Figure ~\ref{mec}
shows a schematic diagram
of the weak charge density.  In the interior region the density is more or less
constant.  In this region, MEC have very little effect.  The density is simply
the conserved weak charge divided by the volume.  It does not matter if the weak
charge resides on the nucleons or on mesons going between the nucleons.  The
only effect of MEC is to slightly change the surface thickness.  This is
indicated by the dotted line in Fig. ~\ref{mec}.  
This change in surface thickness will
only lead to a very small change in the weak radius.  We conclude that MEC are
unlikely to be an issue for the interpretation of the weak radius.

\subsection{Dispersion corrections}

By dispersion corrections we mean multiple electromagnetic or weak interactions
where the nucleus is excited from the ground state in at least one intermediate
state.  At the low momentum transfers considered here, the elastic cross section
involves a coherent sum over the $Z$ protons and is of order $Z^2$.  In
contrast, the incoherent sum of all inelastic transitions is only of order $Z$.
Therefore we expect dispersion corrections to be of order $\alpha/Z$.  This is
negligible.

\subsection{Shape dependence and Surface Thickness}

In principle, the weak radius follows from the derivative of a form factor
evaluated at zero $Q^2$.  In practice, the measurement will be carried out at a
small but nonzero $Q^2$.  Thus the extraction of the weak radius from the
measured form factor may depend slightly on the assumed surface thickness.

We emphasize, one primary use of a measurement is to calibrate mean field models
of neutron densities.  One can simply calculate the weak form factor, 
Eq.~\ref{equation_fwq}, and directly 
compare theory and experiment without any model dependence
or the need to extract a neutron radius.  However, if one wishes to extract a
neutron radius one must address the dependence of the radius on the shape of the
neutron distribution.  One is most sensitive to the surface thickness.

For example, if the weak density of ${}^{208}$Pb is modeled as a Wood Saxon with
radius parameter $c$, 
\begin{equation}
\rho_W(r) = \rho_0 /\bigl\{{\rm
exp}[(r-c)/z]+1\bigr\},
\label{equation_woodsax}
\end{equation}
then the surface thickness parameter
$z\approx 0.55$ fm must be known to $\pm 0.14$ fm in order to extract $R_W$ to
1\% from an asymmetry measurement at the proposed $Q^2=0.008$ GeV$^2$.  Thus the
surface thickness or $z$ must be known to only 25 \% in order to extract $R_n$.

We believe the surface thickness of the weak density is known to much better
then 25\% for at least two reasons.  First, the surface thickness is strongly
constrained by the known surface and single nucleon separation energies.  At
large $r$ the weak density is dominated by the most weakly bound neutron.  This
decays exponentially with a known separation energy.  Therefore the large $r$
behavior of the weak density is known.  At somewhat smaller radii the density is
controlled by the surface energy.  The very abrupt change in density, necessary
for a small surface thickness, implies a very high surface energy.  Any model
with a very small surface thickness will fail to reproduce the known binding
energies of a range of nuclei.  As a result, all mean field models, that we are
aware of, have only a small spread in surface thickness --much less then 25\%--
if they reproduce binding energies .

Second, the surface thickness of the weak density is constrained by the measured
surface thickness of the charge density.  Perhaps the easiest way to change the
surface thickness of the neutron density is to change the thickness of both the
protons and neutrons.  However, this will quickly conflict with the measured
charge density.  Therefore, one has to try and change the surface thickness of
the neutrons without changing the proton density.  This will necessitate large
separations in both energy and position between protons and neutrons.  To
accomplish this, one will need energetic isovector interactions which in turn
will change the binding energies of nuclei as a function of N or Z and ruin
agreement with known masses.

Note, present mean field models do an excellent job reproducing the surface
thickness of measured charge densities.  This is a nontrivial check.  Although
one or more parameters of mean field forces are often fit to charge radii, the
detailed form of the surface density is not fit.  Therefore the excellent
agreement between theory and experiment in the surface region demonstrates both
the power and basic correctness of these arguments that the surface thickness
is strongly constrained by measured binding energies.

\begin{figure}
\vbox to 5.in{\vss\hbox to 8in{\hss {\includegraphics{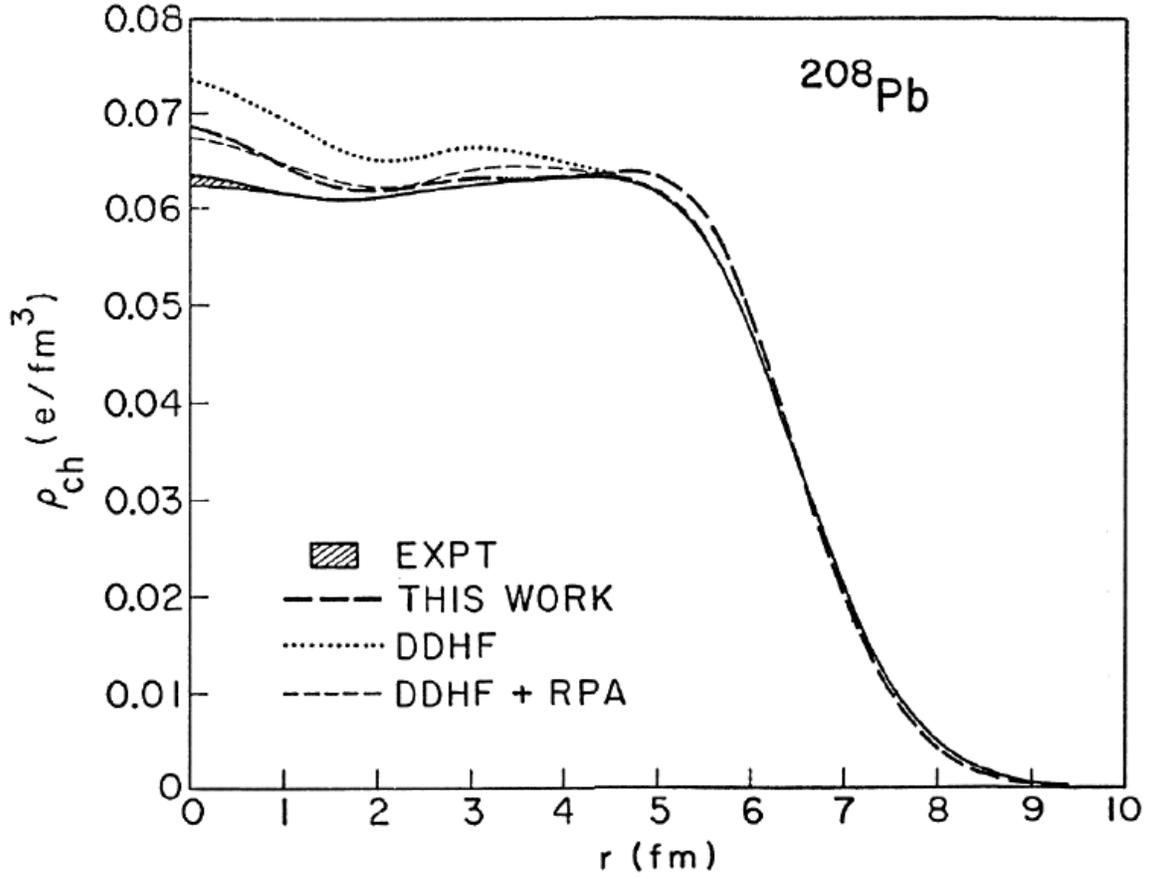}}\hss}} 
\caption{
The charge density in ${}^{208}$Pb.  The
dotted and thin dashed curves are mean field calculations with reasonable
surface energies while the thick dashed curve is a mean field calculation with a
very high surface energy \protect\cite{BDSCJH}.  The experimental charge density is the solid curve.  }
\label{surface}
\end{figure}

We illustrate the above points in Fig.  ~\ref{surface}
This shows the charge density in
${}^{208}$Pb.  A figure for the weak density would be similar.  Conventional
mean field models, thin dashed and dotted curves, agree very well with the
measured surface thickness (region beyond r=5 fm).  (We note that a low $Q^2$
measurement is insensitive to the interior density.)  In contrast the thick
dashed curve shows a relativistic mean field model with a very incorrect surface
energy\cite{BDSCJH}.  This calculation has an incompressibility (which is
closely related to the surface energy) more then a factor of 
two too large.  This
error is well outside of present uncertainties.  Therefore the surface
properties of this calculation can be ruled out.  Nevertheless even with this
large error, the surface thickness disagrees with data and other calculations by
only 10\%.  Since this is less then 25\% there would be no problem using this
incorrect surface to extract the neutron radius to 1\%.

We state the results of this section in a slightly different language.  This
measurement is sensitive to the surface thickness at only the 25\% level, while
it is sensitive to the radius at the 1\% level.  Since 25\% is much larger then
the present spread in surface thickness of mean field models one will not learn
new information about the surface.  Instead a 1\% constraint on the radius does
provide important new information on the radius because present models have a
larger spread then 1\%.

Finally, uncertainties from the surface thickness are even less important in
extracting weak charge information for atomic parity experiments.  This is
because the atomic experiments depend on the surface thickness in somewhat
similar ways to the electron scattering asymmetry.  As a result, some of the
error from the unknown surface thickness cancels in comparing the two
experiments.  Therefore, one could tolerate an uncertainty in the surface
thickness {\it of more then 25\%} and still interpret the atomic experiment.
This is discussed in section VI.

We don't believe the dependence on the surface thickness is a problem.
Nevertheless, if one wanted to reduce the sensitivity there are two options.
First, measure at a lower $Q^2$.  Unfortunately this reduces the magnitude of
the asymmetry and its sensitivity to the neutron radius.  More beam time will be
required and one may be more sensitive to systematic errors.  Alternatively, one
could measure a second asymmetry at a higher $Q^2$.  Within a given model of the
shape of the weak charge density this second point provides information on the
surface thickness.  For example if one assumes a Wood Saxon neutron density,
changing the surface thickness $z$ from 0.5 fm to 0.6 fm (at fixed mean square
radius) decreases the asymmetry by 8\% at 850 MeV and 12 degrees while the
asymmetry is decreased by only 1.5\% at 6 degrees.  Thus the large angle point
is much more sensitive to the surface thickness and in principle could help
constrain it.  However, this second point will require considerable extra beam
time.  Furthermore, the high $Q^2$ point is sensitive to other features of the
shape in addition to the surface thickness.  Therefore, the interpretation of
the high $Q^2$ point may be model dependent.

In summary, one only needs very mild information about the shape of the weak
charge density to extract a radius from the measured asymmetry.  One needs to
know the surface thickness to about 25\%.  We believe this is well within the
accuracy of present mean field models.  We emphasize, even this small ambiguity
does not effect the direct comparison of the measured form factor to theoretical
models.

\subsection{Inelastic Contributions}

In principle, one could measure with enough energy resolution to avoid excited
state contributions.  However in practice, there may be a gain in rate by
running with lower resolution and allowing a small contamination from excited
states.  For example, one can use a thicker target with a larger energy loss.
This contamination is expected to be small because inelastic cross sections at
low momentum transfer are typically much smaller then the elastic cross
sections.

It is useful to estimate the inelastic asymmetry.  The first excited state in
${}^{208}$Pb is at 2.6 MeV and has spin and parity $3^-$.  This is a collective
density oscillation\cite{BM}.  We expect the longitudinal to dominate over the
transverse or axial responses (at forward angles).  In plane wave Born
approximation the asymmetry for a natural parity spin $J$ excitation is
then\cite{donnelly}, 
\begin{equation}
A={GQ^2\over 4\pi\alpha \sqrt{2}} \Bigl\{4{\rm
sin}^2\Theta -1 +{F_n^J(Q^2) \over F_p^J(Q^2)}\Bigr\},
\label{equation_apbinel}
\end{equation}
with $G$ the
Fermi constant.  
Here the neutron transition form factor is, 
\begin{equation}
F^J_n(Q^2)=N\int
r^2 dr j_J(qr) \rho_n^{tr}(r),
\label{equation_fjn}
\end{equation}
in terms of the neutron transition
density $\rho_n^{tr}(r)$ and a similar expression for 
the proton transition form
factor $F_p^J(Q^2)$ in terms of the proton transition density $\rho_p^{tr}(r)$.

The collective density oscillation can be modeled as a deformation of the ground
state density\cite{BM}.  If the elastic neutron density is characterized by a
radius $R_n^0$ then the excited state has a density parameter $R_n^0(\theta)$,
\begin{equation}
R_n^0(\theta) \approx R_n^0[1+\alpha_J^n Y_{J0}(\theta)],
\label{equation_rn0}
\end{equation}
where the small amplitude $\alpha_J^n$ can 
be adjusted to reproduce the magnitude of
the cross section.  Likewise the proton density is characterized by
$R_p^0(\theta)$, 
\begin{equation}
R_p^0(\theta) \approx R_p^0[1+\alpha_J^p
Y_{J0}(\theta)],
\label{equation_rp0}
\end{equation}
with amplitude $\alpha_J^p$.  We assume the radius
parameter $R_n^0$ is proportional to the root mean square radius $R_n$ and
$R_p^0$ is proportional to $R_p$, see 
Eqs.~\ref{equation_r2p}-~\ref{equation_r2n}.

The transition density is then, 
\begin{equation}
\rho_n^{tr}(r)\approx -\alpha_J^n R_n^0
{d\over d r} \rho_n(r).
\label{equation_rhontr}
\end{equation}
The experiment is at a low $Q^2$ well below
the maximum in the inelastic form factor so one can expand the spherical Bessel
function and integrate by parts to obtain, 
\begin{equation}
{F_n^J(Q^2)\over F_p^J(Q^2)}\approx
{\alpha_J^nN \over \alpha_J^p Z }\Bigl({R_n \over R_p}\Bigr)^J.  
\label{equation_fnjq}
\end{equation}
The $3^-$ state has the neutrons and protons oscillating primarily in phase
(``isoscalar''), 
\begin{equation}
\alpha_J^n \approx \alpha_J^p.
\label{equation_alphajn}
\end{equation}
We will discuss
this in more detail below.  With Eq.~\ref{equation_alphajn}
the asymmetry is, 
\begin{equation}
A\approx
{GQ^2\over 4\pi\alpha \sqrt{2}} \Bigl\{4{\rm sin}^2\Theta -1 +{N \over Z}
\Bigl({R_n\over R_p}\Bigr)^J \Bigr\}.
\label{equation_apbinel2}
\end{equation}
In the limit $R_n\approx R_p$
this reduces to, 
\begin{equation}
A\approx {GQ^2\over 4\pi\alpha \sqrt{2}} \Bigl\{4{\rm
sin}^2\Theta -1 +{N \over Z} \Bigr\}.
\label{equation_apbinel3}
\end{equation}
In the same limits, plane
wave and $R_n\approx R_p$, the elastic asymmetry also reduces 
to Eq.~\ref{equation_apbinel3}.
Therefore, {\it the asymmetry for collective natural parity ``isoscalar''
excited states is similar to the elastic asymmetry.}  
This reduces the effect of
the inelastic contamination.

Collective ``isovector'' excitations where the neutrons oscillate out of phase
from the protons, 
\begin{equation}
\alpha_J^p \approx - \alpha_J^n,
\label{equation_jpapprox}
\end{equation}
have a different asymmetry.  
In principle, these could be a concern.  However, we
believe it is possible to use existing $(e,e')$ and $(p,p')$, $(p,n)$, etc.
cross section data to rule out large ``isovector'' strength.

This subsection makes plane wave estimates of the asymmetry.  Unfortunately,
there are at present no distorted wave calculations of the inelastic asymmetry.
We expect coulomb distortion effects to be similar for inelastic and elastic
scattering because the electron wave functions are the same and the plane wave
asymmetries are comparable.  Therefore, our final estimate of the asymmetry of
the $3^-$ state in ${}^{208}$Pb is, 
\begin{equation}
A=D{GQ^2\over 4\pi\alpha \sqrt{2}}
\Bigl\{4{\rm sin}^2\Theta -1 +{N \over Z} \Bigl({R_n\over R_p}\Bigr)^3
\Bigr\}.
\label{equation_apbinel4}
\end{equation}

Here the correction factor for Coulomb distortions $D$ is
taken from the elastic calculations of ref.\cite{cjh}.  At 850 MeV and 6 degrees
this is, 
\begin{equation}
D\approx 0.74\pm 0.26.
\label{equation_dpb}
\end{equation}
We arbitrarily assigned a 100\% error to the 26\% reduction from distortions because there is no explicit inelastic calculation.  Evaluating 
Eq.~\ref{equation_apbinel4}
for a realistic $R_n\approx R_p + 0.2$ fm yields, 
\begin{equation}
A(3^-) \approx 1.25 A ({\rm
elastic}).
\label{equation_3minus}
\end{equation}
This 25 \% enhancement can be understood as follows.
The neutron elastic form factor is reduced because $R_n>R_p$.  Therefore the
elastic asymmetry is about 10\% reduced from Eq.~\ref{equation_apbinel3}
\footnote{In plane
wave}.  In contrast $R_n>R_p$ means that the inelastic form factor will peak at
a lower $Q^2$ then the proton form factor.  Thus the $Q^2$ of the measurement is
slightly closer to the neutron peak then the proton.  As a result $A(3^-)$ is
increased from Eq.~\ref{equation_apbinel3}.

At $q=0.45$fm$^{-1}$ Eq.~\ref{equation_apbinel4} yields,
\begin{equation}
A(3^-)\approx 0.83 \pm 0.29 \pm 0.03 \times 10^{-6}.
\end{equation}
Here the first error is from the distortions and the second error assumes
$R_n$ is known to 1\%.  The 0.83 magnitude is based on arbitrarily assuming
$R_p=5.5$ and $R_n=5.7$fm.

In summary, the asymmetry for the first excited state in ${}^{208}$Pb is
qualitatively similar to the elastic asymmetry.  This reduces the size of the inelastic
correction.  There is some uncertainty
because no inelastic distorted wave calculations have been done.  However,
inelastic contaminations are likely to be small, less then 1\% of the rate, because the elastic cross section at low $Q^2$ is large.  Therefore, the interpretation of a measurement is unlikely to be a problem even if one assumes very large errors for the inelastic $A$.

\subsection{Isospin Violation}

One uses assumptions about isospin symmetry to go from various quark weak
currents to nucleon and eventually nucleus weak current matrix elements.  First,
our formalism, and that used by most others, assumes good isospin in the
nucleon.  For example, Eqs.~\ref{equation_gpz}-~\ref{equation_gnz} 
assume an up quark matrix element in the
proton is the same as a down quark matrix element in the neutron.  This is, no
doubt, violated at some level.  However, calculations such as ref.\cite{Pollock}
suggest only very small corrections.  Thus we don't expect our results to be
impacted by isospin violation in the nucleon.

Second there is, of course, isospin violation in a heavy nucleus.  For example
the proton radius is different from the neutron radius.  In a light or medium
mass nucleus it is often convenient to use an isospin formalism.  This might start
out with equal proton and neutron radii.  In this case, one must explicitly
include corrections to the asymmetry from isopin violation.

In contrast isospin symmetry is not very good for a heavy nucleus with $N>Z$.
Therefore, in the present paper we use a formalism which treats protons and
neutrons separately and does not assume good nuclear isospin.  We simply allow
the proton and neutron densities to be independent.  The weak charge density is
calculated in Eq.~\ref{equation_rhow}
by separately adding proton and neutron contributions.
As a result, we do not need to include any further corrections for isospin
violation in the nucleus.

Note, corrections to the assumption that the weak charge density is the sum of
proton and neutron contributions, for example from density dependent form
factors, can be considered meson exchange currents.  Meson exchange currents
have been discussed in section V.D and 
are expected to be small on quite general grounds.

\subsection{Target Impurities}

A practical experiment could use a backing material to help support and cool the
target.  This could allow a higher beam current and reduce the required beam 
time.  However, such an impurity may complicate the interpretation of the
experiment.  The parity violating asymmetry of the impurity may not be known
exactly.  In addition, the impurity may introduce additional problems such as
low lying excited states.  In this section we discuss one possible composite
target.

It may be possible to make a $^{208}$Pb target with one or more Pb foils 
sandwiched between thin diamond foils.  This may also work for Ba.  The large 
Pb cross section will ensure that only a small fraction of the counts, less 
then say 5\%, are from Carbon.  If such a sandwich is feasible, it may be an 
elegant solution for several reasons.  First the high thermal conductivity of 
the diamond will efficiently transfer the beam power and keep the target from 
melting.  Second, $^{12}$C has a very high first excited state, above 4 MeV, 
so one should not have to worry about excited states.  Finally, $^{12}$C is a 
light nucleus with $N=Z$ so the parity violating asymmetry is very simple and 
well known with only very small uncertainties.  Indeed, $^{12}$C has been used 
for a Standard Model test where it was assumed the asymmetry is so well known 
that any deviation tests the Standard Model\cite{carbon12}.

We have calculated the parity violating asymmetry for $^{12}$C at the proposed
Pb kinematics (850 MeV and six degrees), 
\begin{equation}
A(^{12}{\rm C})=0.660 \times
10^{-6},
\label{equation_carbon}
\end{equation}
this includes a 0.4\% increase from Coulomb distortions
and another 0.4\% increase from differences between the neutron and proton radii
in Carbon.  Equation ~\ref{equation_carbon}
used relativistic mean field densities\cite{BDSCJH}
where the proton radius is slightly larger because of Coulomb interactions.  The
Coulomb distortion correction is both small and has a very small uncertainty.
The uncertainty in the neutron radius correction is larger.  Nevertheless, it is
very unlikely to be orders of magnitude larger then 0.4\%.  Note, a 0.4\% error
in a 5\% impurity is over two orders of magnitude away from impacting a 3\%
asymmetry measurement.  We conclude that the uncertainty in the asymmetry of a
$^{12}$C backing is two orders of magnitude smaller then needed.  Thus such a
backing should pose no problems to the interpretation of an experiment.

This theoretical argument that the carbon backing does not pose a problem is
very strong.  In the remote chance that there is still a question we note that
the asymmetry of ${}^{12}$C has been directly measured in an older Bates
experiment\cite{carbon12}.  This measurement has a somewhat crude error of 20 to
25 \% and is at a somewhat higher $Q^2$.  However, on very general grounds one
expects the asymmetry to scale approximately with $Q^2$.  Therefore it is
straight forward to extrapolate the Bates measurement to our $Q^2$ with an
extrapolation error that is probably significantly smaller then the experimental
error.  We note that even a 25\% measurement of a 5\% impurity only contributes
about 1\% and this is smaller then the goal of 3\% for a ${}^{208}$Pb
measurement.

If one considers drastic deviations of the ${}^{12}$C neutron density then the
Bates experiment should be more sensitive because its higher $q$ is closer to
one over the ${}^{12}$C radius.  A neutron experiment in ${}^{208}$Pb will be
run at a $Q^2$ which is much less sensitive to the ${}^{12}$C density because Pb
is much bigger then C.  Therefore the Bates measurement can help rule out even
theoretically unimaginable ${}^{12}$C neutron densities and insure that a carbon
backing is not a problem.

\subsection{Asymmetry Correction Conclusions}

In this section we have tried to discuss all known corrections to the parity
violating asymmetry.  We find that an experiment determines a well defined form
factor of the weak charge density that can be directly compared to theoretical
models.  A point neutron density can be determined from the measured weak charge
density.  Finally, the dependence on the surface thickness is small so one can 
extract a neutron root mean square radius to 1\% from a single moderately 
low $Q^2$ point.

On a practical side, contributions of excited states or target impurities should
not pose a threat to the interpretation of a measurement.  Therefore an
experiment could be run allowing some inelastic contamination and using a target
backing.  This should reduce the beam time needed.

\begin{figure}
\vbox to 7.5in{\vss\hbox to 8in{\hss {\includegraphics{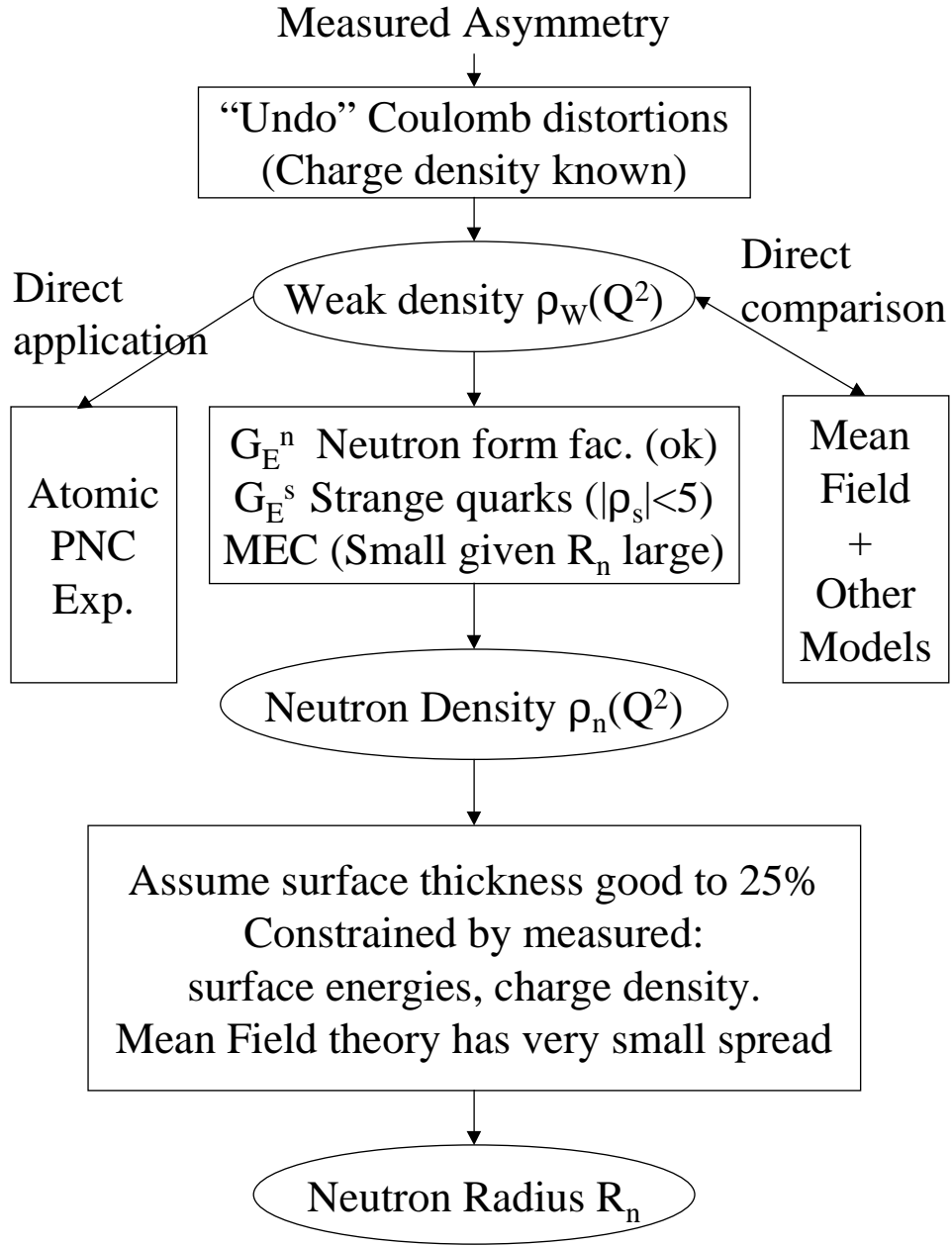}}\hss}}\nobreak 
\caption{
Flowchart of the physics data analysis of
a neutron radius experiment, see text.}
\label{flowchart}
\end{figure}

The physics data analysis of an experiment 
is summarized in Fig. ~\ref{flowchart}.
From the measured asymmetry one can deduce the weak form factor.  This is the Fourier
transform of the weak charge density at the momentum transfer of the experiment.
To deduce the weak density one must correct for Coulomb distortions.
This can be done accurately because the charge density is known. There 
are now a number of independent Coulomb distortion codes.

The weak charge density can be directly compared to predictions of mean field or 
other theoretical calculations.  This will allow isovector interactions to be 
constrained.  The weak charge density can also be applied to atomic PNC 
experiments.  As discussed in the next section this application to atomic PNC 
is almost insensitive to the neutron density surface.

From the measured weak charge density one can deduce a point neutron density by
making small corrections for known nucleon form factors.  The uncertainty in 
these corrections from strange quarks, the neutron electric form factor and MEC
is small.

Finally from this low $Q^2$ measurement of the point neutron density one can
deduce $R_n$.  Because the measurement is at low but not zero $Q^2$ one needs
some very mild information on the shape of the neutron density.  The surface
thickness must be known to about 25\% to extract $R_n$ to 1\%.  All reasonable 
mean field models have a spread in surface thickness much less then 25\%.  
Therefore any mean field shape can be used to extract an equivalent 
$R_n$.

The physics results of the experiment are the weak charge density, the point 
neutron density and $R_n$.  The single number $R_n$ accurately summarizes the
other information.  However, if there is ever a question about the very mild
assumptions on the surface thickness, or if one wishes to consider truly 
drastic changes in the surface thickness which are well outside the range of 
present theory then one can use the neutron or weak density information rather
then $R_n$.

\section{Connection to Atomic Parity Nonconservation}

Parity violating electron scattering (PVES) measurements of the neutron
density will have an important impact on atomic parity nonconservation
(PNC) measurements.  In the future, the most precise low energy tests
of the Standard Model will require a combined knowledge of neutron
densities and atomic PNC observables.  In this section we discuss the
quantitative relationship between PVES and atomic PNC.  As an
instructive illustration, we use approximate parametrizations of the
neutron density and calculate the relative sensitivity of both PVES and
atomic PNC to the shape of the nuclear distribution.  For both PVES and
atomic PNC, these simplified analytical approximations agree well with
the more precise numerical solutions\cite{cjh,Pollock}.  The
sensitivity to the neutron distribution shape parameters is found to be
approximately the same for PVES and atomic PNC, at the kinematics where
PVES is most feasible.

Atomic parity violation experiments can measure the weak charge of a
nucleus\cite{Wieman99}, which at tree level in the Standard Model is
$Q_W^{St.  Mod}=(1-4\sin^2\theta_W)Z-N$.  The effect of finite nuclear
extent is to modify N and Z to $q_n N$ and $q_p Z$
respectively\cite{Pollock}, where

\begin{equation} 
q_{{n} ({p})} = \int f(r) \rho_{{n} ({p})}(r) d^{3} r .  
\label{weakcharge}
\end{equation} 
This nuclear structure correction involves an overlap integral somewhat
similar to the weak form factor of Eq.~\ref{equation_fwq}, 
but here $f(r)$ is a
$q$-independent folding function determined from the radial dependence of
the electron axial transition matrix element inside the nucleus.  If
the opposite parity atomic states which mix are labelled s and p then
$f(r) \propto \psi_p^\dagger(r)\gamma_5\psi_s(r)$.  We avoid computing
the absolute normalization of the electronic wavefunctions, a
calculation requiring full many-body atomic wave function correlations,
by setting f(0)=1.  The approximations we have already made are as
follows:  We treat the nucleons nonrelativistically, ignoring weak
nuclear magnetism effects, and we neglect non-nucleonic degrees of
freedom. We neglect terms involving the vector-electron interaction,
and thus axial- or anapole- nuclear interactions (Experimentally, this
requires properly averaging over hyperfine transitions).  The required
electron s- and p- wave functions can be computed by solving the single
electron Dirac Equation in the presence of the nuclear charge
distribution. By doing this, we neglect effects of electron shielding
in the vicinity of the nucleus, and the effects of electronic binding
energies (and thus many-body correlations) because these are small in
comparison with the nuclear Coulomb potential at short (fm) distances.
level

Atomic theorists make predictions for atomic observables including a
complete many body computation of the axial matrix elements with proper
norm\cite{atomic1}.  To date, they have generally assumed an
isoscalar nuclear density distribution $\rho_n(r)=\rho_p(r)$, and
factored the effect of finite nuclear size into a coefficient of $Q_W$.
The fact that $\rho_n \ne \rho_p$ means that there will then be a small
additive correction which should be made:
\begin{equation}
Q_{{w}}^{\rm expt}=Q_{{w}}^{\rm St. Mod} + \Delta
Q_{{w}}^{{n-p}},
\label{weakchargeii}
\end{equation}
where $Q_w^{\rm expt}$ is the weak charge extracted from atomic
experiments, using atomic theory calculations which {\it ignore} any
neutron-proton differences, and to a good approximation,
\begin{equation}
\Delta Q_{{w}}^{{n-p}} = N(1-q_{{n}}/q_{{p}}).
\label{chargeshift}
\end{equation}
$\Delta Q_W^{n-p}$ is zero if neutron and proton distributions are
identical.  There are additional small 
corrections\cite{Marciano} to $\Delta Q_{{w}}^{{n-p}}$ arising from
Standard Model radiative corrections, as well as additive corrections
arising from e.g. internal structure of the nucleon, but these can be
safely neglected since $\Delta Q_{{w}}^{{n-p}}$ is itself so small.
Given nuclear structure model predictions for $\rho_p(r)$ and
$\rho_n(r)$, the calculation of $f(r)$ and $\Delta Q_w^{n-p}$ is
reasonably straightforward, requiring a numerical solution of the Dirac
Equation in the vicinity of the nucleus. Results from various nuclear
structure model distributions for several nuclei are given in 
Table~\ref{sjptabi},
which shows that neutron-proton distribution differences can affect
measurements of the weak charge at marginally measurable levels. The
model dependent uncertainty in $\Delta Q_w^{n-p}$ appears to be
comparable to the value itself, and exceeds Standard Model radiative
correction uncertainties.  
This motivates improved knowledge of the
neutron distributions, since charge distributions are generally well
measured experimentally.

For the specific case of atomic Cs, recent measurements of transition
polarizabilities~\cite{Wieman99}, coupled with previous measurements of
parity nonconservation (PNC)~\cite{Wieman98} have significantly reduced
uncertainties associated with the extraction of $Q_{{w}}$(Cs).  The
latest result\cite{Wieman99}, $Q_{{w}}^{\rm expt} = -72.06 (.28)_{\rm
expt}\ (.34)_{\rm atomic\ theory}$ is in mild disagreement, at the
$2.5\sigma$ level, with the Standard Model prediction of $Q_{{w}}^{\rm
St. Mod.} = -73.20 (.13)_{\rm theory}$.~\cite{Marciano} The experimental
number uses input from atomic theory calculations~\cite{Johnson,Dzuba}
which incorporate finite nuclear size effects, but do not include the
modification due to neutron-proton differences, $\Delta Q_w^{n-p}$.
Using a relatively naive approximation described below we find that, in
order to reduce the contribution of nuclear structure uncertainty to
below the level of present {\it atomic} theory levels, one would need
to know the neutron radius to around $\pm$ 6\%. To reduce nuclear
structure uncertainties well below the level ($\pm 0.13$) of Standard
Model radiative correction uncertainties requires knowledge of the
neutron radius to around $\pm 2\%$. Due to the more complex nature of
the Cs nucleus (e.g. it is not spin 0) it is unlikely that a PVES
experiment will measure the neutron radius of Cs directly, but
measurement in a nearby nucleus with comparable N/Z (such as Ba) should
provide significantly improved confidence in the ability of nuclear
structure models to predict neutron radii in general, and hence reduce
the nuclear model dependence of these important Standard Model tests.

Precise numerical codes have computed the effect of the neutron density
on PVES\cite{cjh}.  It is instructive to recover this result with a
simpler analytical approximation, and then to connect this to atomic PNC.
To this end we first approximate $f(r)$ and $\Delta Q_w^{n-p}$ by assuming a
uniform nuclear distribution, i.e.  $\rho(r)$ a constant out to some
radius $C$. In this case, the nuclear potential energy is just
\begin{equation}
V(r)=(Z\alpha)*\cases{(-3 + r^2/C_p^2)/2C_p&if  $r<C_p$   \cr
                 -1/r &if  $r>C_p$,\cr}
\end{equation}
neglecting small contributions of neutrons to the nuclear charge
distribution. The single electron Dirac Equation can be solved in
the presence of this potential by expanding in powers of $(Z\alpha)$, 
making a power series for the Dirac wave functions inside the nucleus. 
\footnote{We have also calculated the $(Z\alpha)^4$ corrections, as
well as terms of order $(m_e C_p)(Z\alpha)$ which arise if the electron
mass is left in the Dirac Equation. Including these small corrections,
our approximate formulas reproduce detailed numerical results for $q_n$ at
around the 1\% level.} 
The result is
\begin{equation}
f(r) =  1-(Z\alpha)^2\left(\textstyle{1\over2} (r/C_p)^2 - 
\textstyle{1\over10} (r/C_p)^4 + \textstyle{1\over 150}
(r/C_p)^6\right) + {\cal O}(Z\alpha)^4
\end{equation}
One further simplifying approximation can be made, 
that $\langle r^2\rangle_n \approx
\langle r^2 \rangle_p$, characterizing the difference
by a single small parameter,$\langle r^2\rangle_n/\langle r^2\rangle_p
\equiv 1+\epsilon$.
The result, after applying Eq.~\ref{weakcharge} is
\begin{eqnarray}
q_{{p}} \approx & 1-(Z\alpha)^2(.26),\\
\label{approxi}
q_{{n}} \approx & 1-(Z\alpha)^2(.26+.221\epsilon),\\
\label{approxii}
\Delta Q_{{w}}^{{n-p}} \approx & N (Z\alpha)^2 (.221
\epsilon)/q_{{p}} .
\label{approxiii}
\end{eqnarray}

This can be compared with the 
PVES asymmetry in elastic electron-nucleus scattering for a spin-0
nucleus.  
In the Born approximation (and in the
absence of isospin violation) this asymmetry
is given by Eqn.~\ref{equation_bornasy}
Using the same approximations as
above (uniform distributions, $C_n \approx C_p$) and defining
$A^{\rm nom}\equiv A^0_{lr}(Q^2)\left((1-4\sin^2\theta_W) - N/Z\right)$, we
find
\begin{eqnarray}
{(A-A^{\rm nom})\over A^{\rm nom}} &=&
	\left[ {\displaystyle -N/Z}\over 
		{\displaystyle (1-4\sin^2\theta_w-N/Z)} \right]
 \left({\displaystyle F_n(Q^2)\over {\displaystyle F_p(Q^2)}}-1\right) \\
&\approx&
 \left[1.06\right] \epsilon\left(
	-{3\over 2} + {1\over 2} 
	\displaystyle{(q C_p)^2 \sin(q C_p)\over{	
		(\sin(q C_p) - (q C_p) \cos(q C_p))}}
	\right) + {\cal O}(\epsilon^2)
\label{asymm}
\end{eqnarray}
For extremely small momentum transfer, the expression in large
parentheses can be expanded, yielding $-(q C_p)^2/10 + {\cal O}(q
C_p)^4$.  Knowing $C_n$ to $\pm$1\% means, according to
Eq.~\ref{approxiii} that the weak charge for lead would have an
uncertainty due to neutron structure of $\approx \pm 0.2$, to be
compared with $Q_W^{\rm St. Mod} = -118.7\pm 0.2\ ({\rm rad.\ corr})$.
According to Eq.~\ref{asymm}, at q=0.45 fm$^{-1}$, measuring $C_n$ to
1\% requires an asymmetry measurement with errors around the  3\%
level, in agreement with the numerical results obtained in
ref\cite{cjh}.


The approximation scheme described above can be extended to include
rough effects of the nuclear {\it shape} using the method of
Sandars\cite{Sandars}, adding a ``thin edge" to the uniform
distribution, parameterized by a new skin-thickness parameter $\eta$,
defined for {\it any} arbitrary distribution by
\begin{equation}
\eta
={21\langle r^4\rangle \over {25 \langle r^2\rangle ^2}}-1.
\label{etadef}
\end{equation} 
This form is chosen so $\eta = 0$ for a uniform distribution.
Typically, $\eta_{proton} \approx 0.10$ for a nucleus such as lead.
The presence of a thin edge changes all the moments:
\begin{equation}
\langle r^n \rangle \approx \displaystyle{3\over {n+3}} C^n \left(
       1+\eta \displaystyle{n(n+3)\over8}\right),
\label{edge}
\end{equation}
from which $\langle r^2 \rangle = C^2 (3/5)(1+1.25 \eta)$
serves to {\it define} $C$ for any distribution.  
Adding such a ``thin skin" to the protons, the
charge distribution is unchanged except in a small region near C,
and our
approximation for $f(r)$ is still fairly accurate. The presence of the thin
skin does slightly modify the potential inside the nucleus, 
which in turn modifies
$f(r)$ in a well-defined way.
Adding a thin skin to the
neutrons as well, assuming the
difference $\Delta \eta \equiv \eta_n-\eta_p \ll 1$, 
Eq.~\ref{approxiii}  is then modified to 
\begin{equation}
\Delta Q_{{w}}^{{n-p}} \approx  N (Z\alpha)^2 .22\left(
\epsilon - 0.16                
\Delta \eta \right)/q_{{p}} .
\label{weakii}
\end{equation}
The insensitivity of the atomic observable to higher moments beyond the
rms radius is seen from the small relative coefficient of $\Delta
\eta$.   

To connect to a more familiar measure of skin thickness, consider a
Wood-Saxon form for the density,
with radius and thickness parameters $c$ and $z$, 
as given in equation ~\ref{equation_woodsax}.
An analytic series expansion as
$z/c \rightarrow 0$ gives $\eta \approx \textstyle{4\over3} \pi^2
(z^2/c^2)(1 - \pi^2 z^2/c^2) +...$, $\langle r^2 \rangle =
\textstyle{3\over5}c^2 (1+\textstyle{7\over3}\pi^2 z^2/c^2)$, and $C^2
\approx c^2(1+\textstyle{2\over3}\pi^2 z^2/c^2 + \ldots)$ 
In this way,
we could eliminate $\Delta \eta$ in favor of $\epsilon$ and 
$\Delta z/z$, where e.g. $\Delta z\equiv z_n-z_p$ is a
difference in Wood-Saxon neutron and proton thickness parameters, 
and as stated above, 
$\epsilon = \langle r^2\rangle_n/\langle r^2\rangle_p -1$.
Evaluating $f(r)$
numerically in the presence of this potential 
using the analytic series
expansion for moments and linearizing in $\epsilon$ and $\Delta z$, 
\begin{equation}
\Delta Q_{w}^{n-p}(Pb) \approx 10.6 \epsilon - .37 \Delta z/z,
\label{weakiipb}
\end{equation}
while for barium
\begin{equation}
\Delta Q_{{w}}^{{n-p}}(Ba) \approx  3.3 \epsilon - 0.13  \Delta z/z.
\end{equation}
In these expressions, higher order effects in $(Z\alpha)$, as well as
finite surface thickness, have been taken into account numerically. (We
use as nominal inputs\cite{atomic2}
$c^{Pb} = 6.624$,
$z^{Pb} = 0.549$,
$c^{Ba} = 5.700$,
$z^{Ba} = 0.5314$)


The PVES asymmetry also gets modified by a thin skin, and
Eq.~\ref{asymm} gets an additional 
correction term,
\begin{equation}
{(A-A^{\rm nom})\over A^{\rm nom}} \approx 
 {\rm Eq}.~\ref{asymm} + \left[1.06\right] \Delta\eta \left(
	\displaystyle{15\over8} + (q C_p)^2
	\displaystyle{{q C_p \cos(q C_p)- 6 \sin(q C_p) }\over{	
	       8(\sin(q C_p) - (q C_p) \cos(q C_p))}}
	\right) 
\label{asymmagainii}
\end{equation}
(neglecting any ${\cal O}(\eta_p)$ corrections here.) In the limit of small
momentum transfer, 
the term in large parenthesis goes to $(q C_p)^4/280$.  The PVES
asymmetry thus becomes completely insensitive to $\Delta \eta$ at small
$Q^2$, as one would expect, but at larger $Q^2$ the surface shape becomes
relatively more important.  Table~\ref{sjptabii} shows the ratio of the
coefficients of $\Delta \eta$ to $\epsilon$ in Eq~\ref{asymmagainii} as
a function of momentum transfer. 
(Note that these numbers implicitly incorporate the
approximation $\eta_p \rightarrow 0$.) 
Incorporating finite skin thickness with a Wood-Saxon
for the nucleon distributions, the asymmetry can be calculated
using asymptotic expansion formulas in c and z, and the result
linearized in the small quantities  $\epsilon$ and $\Delta z$. The
resulting formula is not especially illuminating, but the coefficients
of this expansion are shown for lead as a function of momentum transfer
in Table~\ref{sjptabiii}. 

Note that there is a unique momentum transfer where the two observables
are sensitive to 
the {\it same} linear combination of neutron radius and
surface shape.
For lead, assuming a thin edge (Table~\ref{sjptabii}),
and comparing with Eq.~\ref{weakii}, this ``matchup" is around $q=0.34$
fm$^{-1}$.  Using the Wood-Saxon form to incorporate the effects of
finite skin thickness, and comparing with Eq.~\ref{weakiipb}, the relative
coefficients of $\Delta z/z$ and $\epsilon$ are  matched for electron
scattering and atomic PNC at $q\approx 0.32$ fm$^{-1}$.  At this
kinematics point, a measurement of PVES is ``optimized'' to provide the
direct information desired for the atomic observable. In the absence of
other constraints, this would be the optimal momentum transfer for a
PVES measurement if the goal is the most direct measurement of atomic
weak charge corrections, rather than a desire to extract and measure
the details of the neutron shape distribution.  Of course, this is only
true to the extent that still higher moments (shape differences beyond
the simple thin edge approximation) are not important, which is a
decent approximation for the atomic observable, but not so good
for the PVES asymmetry. A brute force estimate obtained by assuming a
generalized three parameter Gaussian form for the nuclear
distributions, allowing the 3 parameters to vary but constraining them
to produce values of A within some small window of $A^{\rm nom}$,  and
then calculating the corresponding spread in $\Delta Q_w^{n-p}$, we
find an optimal q value of $0.32$ fm$^{-1}$.  (All these results,
however, have been calculated at tree level, without Coulomb corrections.)

\begin{figure}
\vbox to 4.in{\vss\hbox to 8in{\hss
{\includegraphics{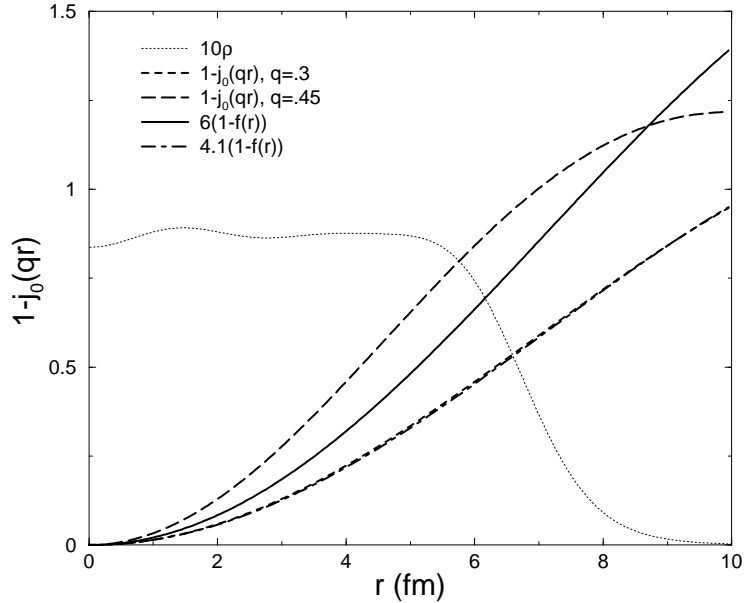}}\hss}}
\nobreak
\caption{Approximate nuclear weak density $\rho(r)$ for Pb, along with
the function multiplying $\rho(r)$ in the integrals for the weak form
factor (namely, $j_0(q r)$) and in the atomic correction factor $q_n$ 
(namely $f(r)$). In both cases the function is subtracted from one 
to eliminate the volume integral of the weak charge density.  
Curves for $1-j_0(q r)$ are shown for two different values of $q$ in fm$^{-1}$.  Note 4.1(1-f(r)) is almost identical to $1-j_0(qr)$ for q=0.30fm$^{-1}$.   }
\label{sinqrfig}
\end{figure}

Another way of understanding the above result 
is to compare the function
multiplying $\rho_n(r)$ in the form factor 
integral of Eq.~\ref{equation_fwq}
(i.e. $j_0(q r)$) and in the convolution 
integral of Eq. ~\ref{weakcharge} (i.e. $f(r)$).
Since we are not interested in the volume integral of the weak charge
density, which is well known, we can subtract both $f(r)$ and
$j_0(q r)$ from 1, and plot the remainder to study the relative
sensitivity to the radius and to the surface thickness.  This is shown
in figure~\ref{sinqrfig}. For $q=0.30$ fm$^{-1}$, $ 1-j_0(q
r)$ is nearly proportional to $1-f(r)$, and thus at this $Q^2$ one is
sensitive to the same ratio of surface and radius in a JLAB parity
experiment and in an atomic experiment. The curves for
$q=0.45$ fm$^{-1}$ are also shown, in this case the curves are not
identical but are similar enough so much of the error from an unknown
surface thickness cancels when comparing the two integrals.

For lead, ignoring the effects of skin thickness, we found above that
the rms neutron radius should be known at roughly the $\pm$1\% level to
ensure that neutron structure uncertainties are smaller than present
Standard Model radiative corrections uncertainties (roughly $\pm$0.2,
or about $\pm$0.16\%, for the weak charge.)  Again ignoring thickness,
table~\ref{sjptabii} shows that this would require e.g. a PVES
asymmetry measurement at q=0.45 fm$^{-1}$ at the $\pm$3\% level.  {\it
Including} the effect of thickness, the linear combination of
$\epsilon$ and $\Delta z/z$ required for the weak charge is not quite
the same as the linear combination measured in PVES at arbitrary $q$, but
a linear error propagation at $q=0.45$ shows that as long as the relative
uncertainty in $z_n$ is less than $\approx \pm$50\%, the additional
uncertainty due to including skin thickness is negligible. Thus, a
single PVES measurement taken even at a kinematics point which is not
perfectly optimized for the atomic observable will still be sufficient
to eliminate nuclear structure effects from the atomic observable at
levels below present Standard Model uncertainties.

\begin{table}
\caption{Nuclear structure model predictions for $\Delta Q_q^{n-p}$.
HFB stands for Hartree Fock Bogolyubov, Skl stands for a Skyrme ``SLy4"
parameter set. G1 is a parameterization from a relativistic nuclear
model. The Standard Model value and uncertainty are from Ref.
\protect\cite{Marciano}.  }
\begin{tabular}{lccc}
Model & Element & $Q_w^{\rm St. Mod}$ & $\Delta Q_w^{n-p}$ \\ \hline
Gogny\protect\cite{decharge}& Pb &  -118.70(0.19) & 0.47 \\ 
HFB-Skl\protect\cite{dob} & `` & `` & 0.57 \\
G1 \protect\cite{horst}& `` & ``  &1.0  \\ \hline
Gogny & Ba &  -77.07(.13) & 0.14  \\
HFB-Skl & `` & ``  & 0.18 \\
G1 \protect\cite{horst}& `` & ``  &0.30  \\ \hline
\label{sjptabi}
\end{tabular}
\end{table}

\begin{table}
\caption{ Table of the coefficients occurring in Eq.~\protect\ref{asymmagainii}, 
written in the form 
$(A-A^{\rm nom})/A^{\rm nom} = c_1(\epsilon + c_2 \Delta\eta)$, 
as a function of momentum transfer, $q$, for a $^{208}$Pb target.
($\epsilon$ measures the n-p rms radius difference,
and $\Delta \eta$ measures the n-p surface shape differences)
Note that Eq.~\protect\ref{asymmagainii} is derived assuming $\eta_p
\rightarrow 0$. 
}
\begin{tabular}{lcc
}
q (fm$^{-1}$)& $c_1$ & $c_2$ 
\\
0.2 	& -.21 	&  -.068 
\\
0.25	& -.33	&  -.11 
\\
0.3 	& -.50	&  -.16 
\\
0.35 	& -.72	&  -.22 
\\
0.4 	& -1.0	&  -.30 
\\
0.45 	& -1.4	&  -.40 
\\
0.5 	& -2.1 	&  -.52 
\\
\label{sjptabii}
\end{tabular}
\end{table}

\begin{table} 
\caption{ Table of the coefficients $a_1$ and $a_2$
occurring in the expression $(A-A^{\rm nom})/A^{\rm nom} = a_1\epsilon +
a_2 \Delta z/z$, as a function of momentum transfer, $q$, for a
$^{208}$Pb target.  The coefficients of $\epsilon$ do not match those
in Table~\protect\ref{sjptabii} for two reasons: finite
thickness {\it is} incorporated here numerically, using the same nominal
inputs as for Eq.~\protect{\ref{weakiipb}}.   Also, note that $\Delta z/z
$ is itself a function of both $\epsilon$ and $\Delta \eta$.}
\begin{tabular}{lccc
}
q (fm$^{-1}$)& $a_1$ & $a_2$ & $a_2/a_1$\\
0.2 	& -.23 	&  .003 & -0.013\\
0.25 	& -.37 	&  .008 & -0.021\\
0.3 	& -.56 	&  .017 & -0.031\\
0.35 	& -.80 	&  .034 & -0.042\\
0.4 	& -1.1 	&  .064 & -0.056\\
0.45 	& -1.6 	&  .12 & -0.073\\
0.5 	& -2.3	&  .21 & -0.094\\
\label{sjptabiii}
\end{tabular}
\end{table}

\section{Conclusion}
With the advent of high quality electron beam facilities
such as CEBAF, experiments for accurately measuring the
weak density in nuclei through parity violating
electron scattering (PVES) are feasible.
The measurements are cleanly interpretable, 
analogous to electromagnetic
scattering for measuring the charge distributions in
elastic scattering.  From parity violating asymmetry measurements
in elastic scattering,
one can extract the weak density in nuclei after
correcting for Coulomb distortions, which have 
been accurately calculated\cite{cjh}.

By a direct comparison to theory, these measurements
test mean field theories and other models that predict 
the size and shape of nuclei.  
They therefore can potentially have a fundamental and lasting
impact on nuclear physics.

Furthermore, PVES measurements have important implications
for atomic parity nonconservation (PNC) experiments which
in the future may become the most precise tests of
the Standard Model at low energies.
We have shown that to a good approximation, sufficient
for testing the Standard Model,
the dependence on nuclear shape parameters enters 
the PVES and PNC observables the same way; 
therefore, the PVES measurements are directly
applicable to the interpretation of atomic PNC if
measured on the same nucleus.

Measurements of the weak density lead to the neutron
density distribution with unprecedented accuracy.
As we have discussed in this paper, PVES yield significant
improvement in the accuracy of neutron densities
compared to hadronic probes or magnetic scattering.
We have shown that the corrections
due to strange quarks, neutron electric form factors, 
parity admixtures, dispersion corrections, 
meson exchange currents, and several other 
possible effects in realistic experiments are all small.
Further, an asymmetry measurement from a 
heavy nucleus with 3\% accuracy
will both establish the existence of the neutron skin
and characterize its thickness.
The neutron skin is an important, qualitative feature of heavy
nuclei which has never been cleanly established in a stable nucleus.

\vskip 2in


\end{document}